\documentclass{article}

\usepackage{diagbox}
\usepackage{svg}

\usepackage[export]{adjustbox}

\usepackage{subfigure}

\usepackage{float}
\usepackage{makecell}
\usepackage{tikz}
\usetikzlibrary{arrows,positioning,decorations.pathreplacing}
\usepackage{url}
\usetikzlibrary{shapes}
\usepackage{xcolor}
\usepackage{balance}
\usepackage{xspace}
\usepackage{amsmath}
\usepackage{hyperref}
\usepackage{bigstrut}
\usepackage{empheq}
\usepackage{cleveref}
\usepackage{amsfonts}% 
\usepackage{multirow}
\usepackage{graphicx}
\usepackage{booktabs}
\usepackage{caption}
\usepackage{tcolorbox}
\usepackage{adjustbox}
\usepackage{array}
\usepackage{numprint}
\usepackage[preprint,nonatbib]{neurips_2022}

\usepackage[utf8]{inputenc} % allow utf-8 input
\usepackage[T1]{fontenc}    % use 8-bit T1 fonts
\usepackage{hyperref}       % hyperlinks
\usepackage{url}            % simple URL typesetting
\usepackage{booktabs}       % professional-quality tables
\usepackage{amsfonts}       % blackboard math symbols
\usepackage{nicefrac}       % compact symbols for 1/2, etc.
\usepackage{microtype}      % microtypography
\usepackage{xcolor}         % colors

\title{FLock: Defending Malicious Behaviors in Federated Learning with Blockchain}

\author{
Nanqing Dong\thanks{Authors are arranged in alphabetical order.}\\
Department of Computer Science\\
University of Oxford\\
\texttt{nanqing.dong@cs.ox.ac.uk}\\
\And
Jiahao Sun\\
FLock.io\\
\texttt{sun@flock.io}\\
\And 
Zhipeng Wang\\
Department of Computing\\
Imperial College London\\
\texttt{zhipeng.wang20@imperial.ac.uk}\\
\And
Shuoying Zhang\\
FLock.io\\
\texttt{shuoying@flock.io}\\
\And
Shuhao Zheng\\
School of Computer Science\\
McGill University\\
\texttt{shuhao.zheng@mail.mcgill.ca}\\
}

\begin{document}

\maketitle

\begin{abstract}

Federated learning (FL) is a promising way to allow multiple data owners (clients) to collaboratively train machine learning models without compromising data privacy.
Yet, existing FL solutions usually rely on a centralized aggregator for model weight aggregation, while assuming clients are honest.
Even if data privacy can still be preserved, the problem of single-point failure and data poisoning attack from malicious clients remains unresolved.
To tackle this challenge, we propose to use distributed ledger technology (DLT) to achieve 
\underline{\textbf{FLock}}, a secure and reliable decentralized \underline{\textbf{F}}ederated \underline{\textbf{L}}earning system built on bl\underline{\textbf{ock}}chain.
To guarantee model quality, we design a novel peer-to-peer (P2P) review and reward/slash mechanism to detect and deter malicious clients, powered by on-chain smart contracts.
The reward/slash mechanism, in addition, serves as incentives for participants to honestly upload and review model parameters in the FLock system.
FLock thus improves the performance and the robustness of FL systems in a fully P2P manner.

\end{abstract}

\section{Introduction}
\label{sec: intro}

% What is blockchain? Why we should use blockchain in FL?/what is the motivation of using blockchain in FL?
% Briefly describe our method and summarize the contributions. 
% For Nature, what is the overall impact?

% What is FL? 
Federated learning (FL)~\cite{mcmahan2017communication} is a machine learning (ML) paradigm where data owners/clients collectively train a ML model under the orchestration of a central server, while the clients' raw data stay local -- they are not transferred or shared with any other party. Most FL implementations follow a centralised, 'hub-and-spoke' topology. The aggregation server distributes global models for clients to train, collects their model updates, then re-distribute updated global model, until training is complete. \cite{li2014communication}
% Why we need FL/what is FL used for?

Since FL does not involve collecting data from all data services and storing it on a server, it is seen as a promising solution to conduct machine learning while protecting user privacy. Privacy-preserving machine learning has become increasingly valuable due to regulatory restrictions and customer demand. Globally, over $70\%$ of countries have introduced regulations on data privacy; in particular, GDPR~\cite{gdpr} in Europe and HIPAA~\cite{hipaa} in the US have strict restrictions on what companies can access and store. This has either increased compliance and operations costs, or resulted in companies discontinuing related services \cite{data-protection-and-privacy}. With FL, companies can analyse data without centralising them first, resulting in simpler compliance procedures and fewer regulatory restrictions. On the other hand, consumers face the dilemma between data protection and functionality. According to a 2021 survey~\cite{a-customer-centric-approach}, while two thirds of the customers do not have a positive view on companies' data protection policies, under $20\%$ are happy to share personal information to get additional value such as better services -- the rest either do not want to share or consider it as a compromise. FL will bring a much stronger user proposition, one that makes data collection obsolete and does not ask users to choose between privacy and functionality.  

% What is the current bottleneck in FL? 

In conventional FL settings, clients submit local model updates to the central server, which stores and aggregates local updates, before sharing global model updates with the clients back. The central server could become a single point of failure for both the security and the performance of the system, prone to model poisoning, privacy leakage, network delay, and targeted delays. 

Blockchain~\cite{wood2014ethereum}, which acts as an immutable and decentralised ledger, is well-placed to tackle the issue of the server being a central point of failure. There have been several proposed approaches replacing a central server in  with a blockchain~\cite{wang2021blockchain}.

% What is the current bottleneck in FL x Blockchain? 
A drawback is that these approaches assume that clients do not intend to undermine the global model and can commit resources to staying online and local training. However, there are several ways malicious clients could impair the global model performance. For instance, they could drop out from training, refuse to upload model updates, train models using low-quality or fake data, or just spam the system with random model updates. 

The lack of systematic protection against malicious clients is a limiting factor to FL's widespread and decentralised adoption, in conventional and on-chain settings. Implementation is thus limited to existing trust circles, where rules of collaboration are more enforceable. To pave the way for large-scale cooperation, we need an automated and enforceable mechanism to ensure execution integrity among participants and protect against malicious attempts. 

We aim to combine blockchain's consensus mechanisms with a decentralised scoring mechanism to preserve performance as well as decentralisation. There is no trusted party to authorise which transactions are valid or organise the participants on the blockchain; instead, any party has the opportunity to propose new blocks and the participants are organised by a pre-agreed definition of the canonical record (chain of transactions with the most computation done on it) and an incentive scheme for building on the canonical record (mining rewards). On top of the blockchain, smart contracts allow the execution of arbitrary code or rules, forming a decentralised and distributed machine with shared states among each participating party.

We propose a secure and reliable decentralised Federated Learning system built on blockchain (``FLock'') to reduce the trust in server/other clients required. In addition to on-chain smart contracts replacing the central server in the model aggregation process, there are two main contributions: a reward/slashing mechanism to ensure model performance, and a P2P model evaluation system. Smart contracts provide transparency and ensure there are no side channel attacks from the central server; the reward/slashing mechanism deters data poisoning attacks; and finally, the P2P model evaluation system safeguards model performance and filters out malicious clients in the multi-round setting.

\section{System Design}

% \begin{figure}[htbp]
% \centering
%   \centering
%   \includegraphics[width=\columnwidth]{sections/figures/}
% \caption{System design of the conventional FL.}
% \label{fig:conventional_FL}
% \end{figure}

\begin{figure}[htbp]
\centering
  \centering
  \includegraphics[width=0.83\textwidth]{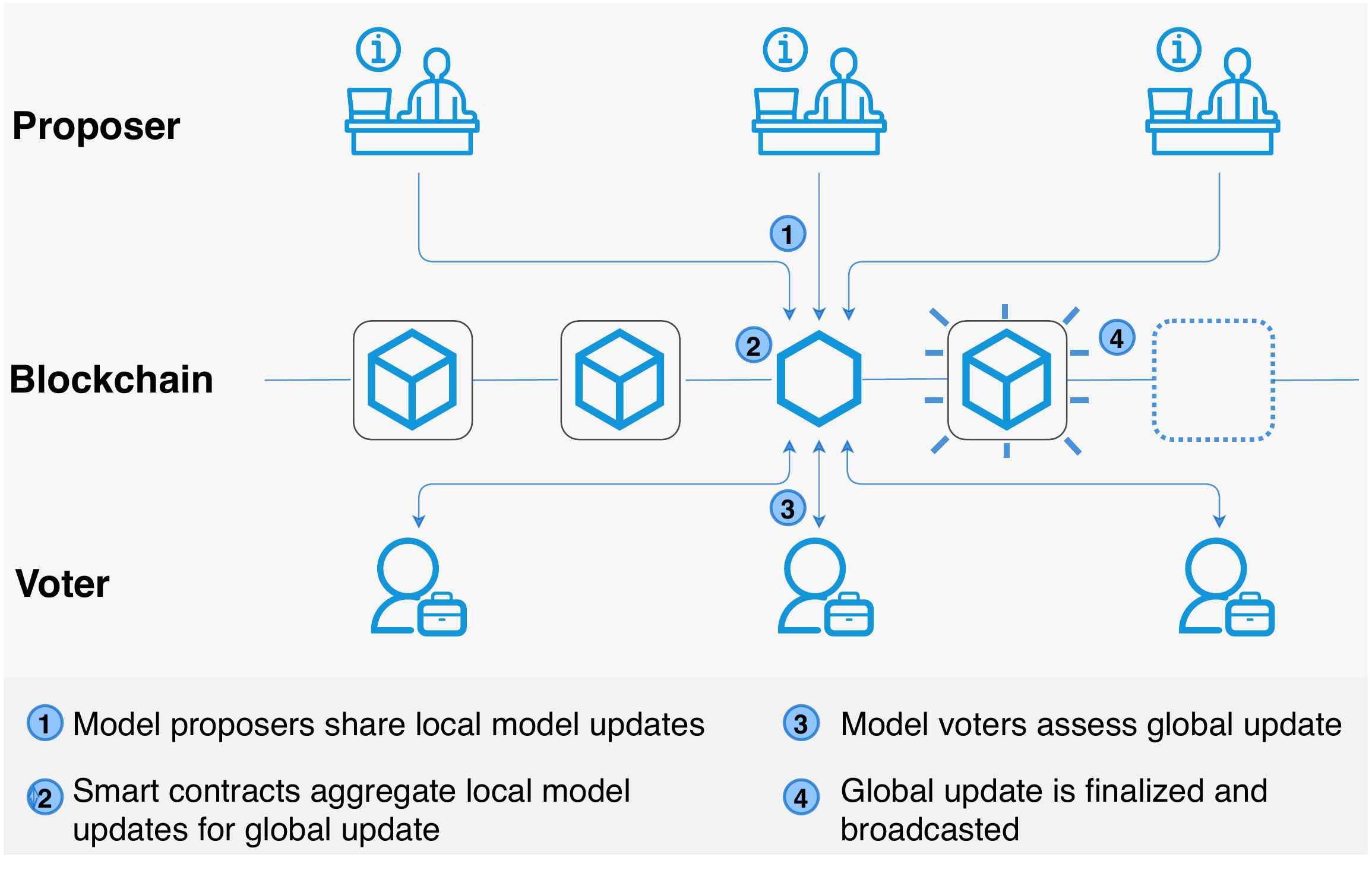}
\caption{System design of Flock FL.}
\label{fig:Flock_FL}
\end{figure}
%Setup Phase

%Training Phase: Local training + On-chain Aggregation + Committee Voting + Reward/Slash

%
The maliciously secure FLock system consists of two phases: namely the \textbf{Setup} phase and the \textbf{Training} phase.
Participants prepare the data and stake tokens during the Setup phase and collaboratively train the model in an auditable way during multiple Training phases, with each Training phase consisting of four steps: \textbf{local training}, \textbf{on-chain aggregation}, \textbf{committee voting}, and \textbf{reward/slash}.

\subsection{Participants}
There are mainly three kinds of participants in FLock system (cf.~Figure~\ref{fig:Flock_FL}):
\begin{itemize}
    \item \textit{Proposer}: Blockchain nodes selected to train the model on their local private dataset and report weight updates.
    \item \textit{Voter}: Blockchain nodes selected to form a committee to audit the aggregated model weights during each Training phase.
    \item \textit{Miner}: Blockchain nodes that actively order transactions on the FLock blockchain and produce blocks.
\end{itemize}

In this work, we mainly focus on describing the unique working procedure for clients and voters, while the miner's part is similar to that in other blockchain systems.
It is worth noting that one person in the real world is not restricted to acting as only one participant in the FLock system, which adds extra security requirements to the whole system since the malicious strategy becomes more complex in this scenario.
The mechanism design of FLock system also targets to defend against such complex malicious behaviors to achieve a practical system ready to use.

\subsection{System Pipeline}

\subsubsection{Setup Phase}
To participate in a federated learning task, any participant should first stake some tokens into the FLock system as collaterals to represent their commitment.
The staked tokens will later be used in the reward/slash phase to reward honest actors and punish malicious actors.
For miners, staked tokens also contribute to the probability to be selected as the block producer and receive block rewards, known as the Proof-of-Stake (PoS) mechanism.

In order to participate in model training, a proposer should prepare a private training dataset and optionally a validation dataset for local model selection.
Any node who wants to become a committee voter should prepare a local test dataset for global model weight auditing.
Obviously, any proposer can leave out part of their training dataset as the test dataset and participate as a voter.
A novel reward/slash mechanism is applied in the FLock system to incentive honest proposers and voters and punish malicious players, which is explained in Section 
\subsubsection{Training Phase}
\label{subsubsec: training}
One training phase consists of the following four steps:
\begin{enumerate}
    \item \textit{Local training}: 
    Each node receives the same global model weights at the end of each training phase.
    Then, at the beginning of a new training phase, several nodes are randomly selected as model proposers for this phase from all the nodes eligible to train models.
    Only the selected nodes are permitted to locally train their model based on the global model and propose the local model updates to the FLock system.
    %
    % The random selection is performed via a Verifiable Random Function (VRF) to minimize the possibility of collusion.
    %
    % Moreover, to avoid model weight plagiarism, the proposers act in a commit-and-reveal manner in the asynchronous proposing environment.  
    \item \textit{On-chain aggregation}: 
    After receiving the local model weight updates from the proposers in this round, the miner selected to produce the block will aggregate the weights via the FedAvg protocol and publish the global model weights on the blockchain.
    However, as the miner might be malicious and censor specific proposers to gain benefits, it is crucial for other miners to validate the correctness of model aggregation.
    Therefore, after the global model weights are proposed, other miners will start a voting round for the validity of the aggregation results before entering the next step.
    \item \textit{Committee voting}: 
    To avoid bad model weight updates and punish malicious proposers, an auditing committee is randomly selected among all the eligible participants to evaluate the global model weights.
    Each committee voter locally evaluates the model performance on the previous prepared test dataset and calculates a voting score based on the performace.
    % and converts the model performance gain to a voting score $score_i\in [-1, 1]$ representing whether and how much the model is getting better or worse.
    %
    Afterwards, each voter simply reports the voting score to the blockchain miners for tallying the votes.
    %
    % After getting all the votes, the blockchain miner 
    % simply chooses the medium score as the final evaluation for the global model weights in this round, denoted as $aggScore$.
    %
    An automatic on-chain protocol will determine if the global model weights in this round will be discarded according to the voting scores.
    \item \textit{Reward/Slash}: 
    %
    % $aggScore$ acts as the audition results for the model training in each round.
    %
    To incentivize honest behaviors and get rid of bad actors, we re-distribute the staked tokens according to the voting scores at the end of each training phase.
    For the proposers, the reward/slash is given based on the voting scores.
    %
    % Specifically, we set a reward threshold $T_p$ and reward/slash all the proposers in this round if the final $aggScore$ is above/below that threshold.
    %
    % Intuitively, although there could be false positives and false negatives, we configure the system such that in expectation honest proposers will finally be rewarded, and malicious proposers will finally be slashed.
    %
    For the voters, we calculate the amount to reward/slash based on the difference between their individual voting score and the final aggregation score and the voting direction.
    %
    % Concretely, if a voter has the same voting direction as the aggregation score, she will get rewards.
    %
    % Otherwise, when the voting direction is different, a voter will only be slashed if the difference between her score and the aggregation score is larger than some threshold $T_v$.
    %
    We configure the system such that eventually honest participants will get rewards and malicious participants will get slashed.
    Since each participant should stake enough tokens to participate, malicious participants will eventually quit the system as their tokens get slashed.
    In this way, the FLock system facilitates a secure and truthful federated learning process without sacrificing the final model performance. 
\end{enumerate}

\subsection{Threat Model}
The malicious behaviors we consider in designing the system include:
\begin{itemize}
    \item \textit{Malicious Proposer}: 
    A proposer may use bad or duplicate data for training, or report bad model weight updates to harm the global model performance.
    She may also decline to upload her model weight updates or be unresponsive.
    \item \textit{Malicious Voter}: 
    A voter may vote extremely in the reverse direction to maximally bias the aggregation result and gain benefits.
    She may also decline to vote or be unresponsive.

    % {\color{red}
    % \item \textit{Malicious Miner}:
    % %
    % A miner may censor specific proposers or voters by not including their model weights or votes.
    % %
    % She may also decline to produce blocks or be unresponsive.}
\end{itemize}

Since one person in the real world can register for multiple nodes in the FLock system, it is also necessary to consider collusion behaviors among different roles in the system.
In this work, we theoretically prove that the system parameters can be configured to defend against all the possible malicious behaviors under the honest majority assumption, which is generally believed to be true in most PoS blockchain systems.
It is worth noting that defending against malicious miners is not considered in this work, which already has effective solutions such as FlashBots~\cite{daian2019flash}.

\section{Theoretical Analysis}
In this section, we provide theoretical analysis on expected returns for proposers and voters, and calculate the optimal system parameters to incentivise good behaviours and penalize malicious participants.

\subsection{Expected Return for Participants}
The expected rewards for clients who participate in proposing are shown in Equation~\ref{eq:E_R_p}. Note that $N$ is the number of participants, $N_p$ is the number of proposers per round, $N_v$ is the number of voters per round, $l_p$ is the ratio of malicious proposers and $l_v$ is the ratio of malicious voters. $\alpha$ and $\beta$ are two system parameters determining the participants' rewards and penalties. $T$ is the threshold number configured by the system.

\begin{equation}\label{eq:E_R_p}
\begin{aligned}
   \mathbb{E}(R^p) &= f(\alpha, \beta, T, N, N_p, N_v, l_p, l_v)
\end{aligned}
\end{equation}

\begin{figure}[t]
\centering
  \centering
  \includegraphics[width=0.85\textwidth]{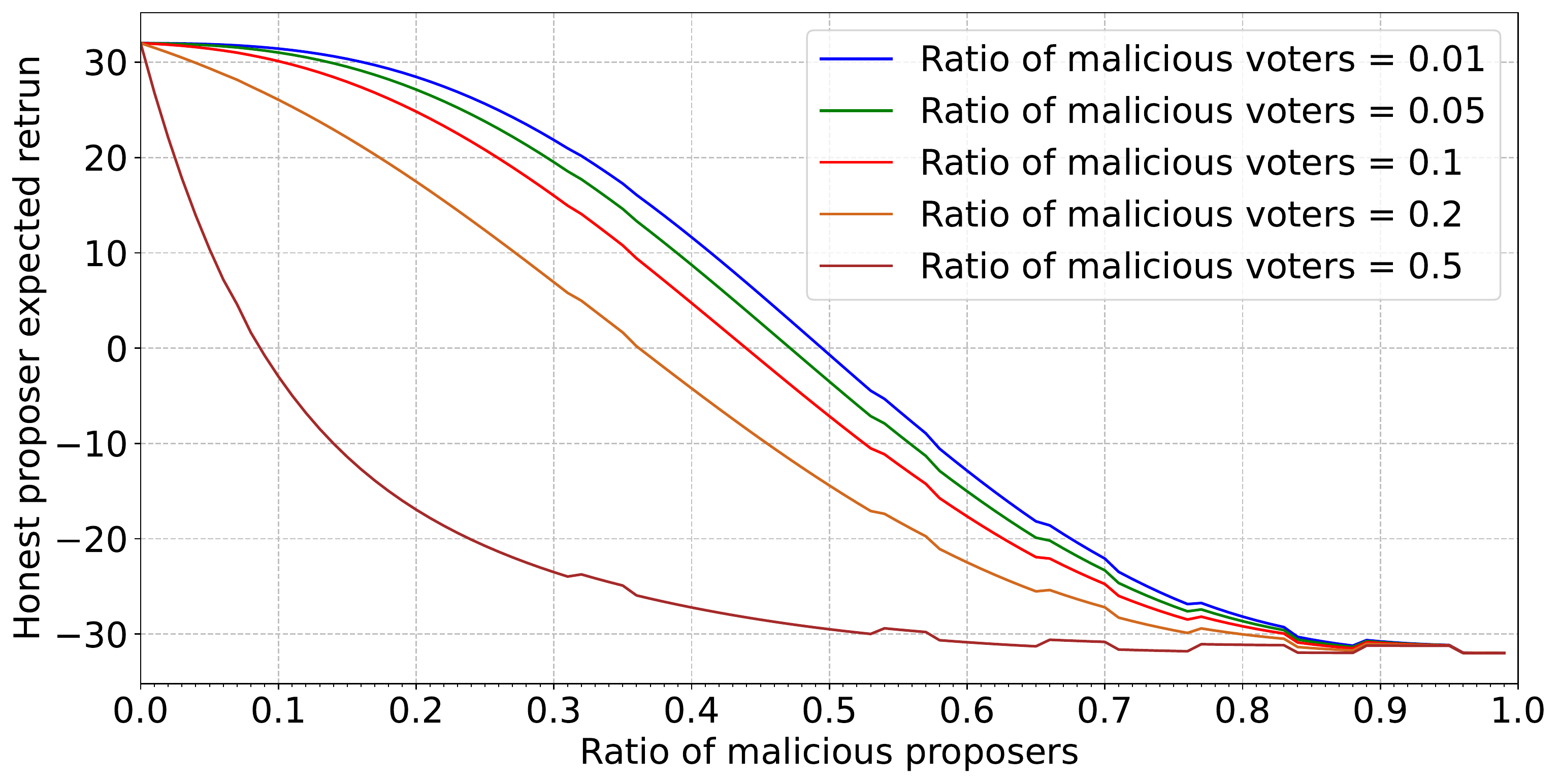}
\caption{The expected rewards of honest proposers over the ratio of malicious proposers.}
\label{fig:ER_honest_over_lp}
\end{figure}

As shown in Figure~\ref{fig:ER_honest_over_lp}, the expected rewards of honest proposers decrease over the ratio of malicious proposers given fixed system parameters. Our theoretical analysis results prove that there are optimal system parameters (i.e., $\alpha$, $\beta$ and $T$) to incentivise honest participants and penalize malicious participants.

% %%%%%%%%%%%%%%%%%%%%%%%%%%%%%%%%%%%%%%%%%%%%%%%%%%%%%%%%%%%%

\begin{ack}
This work was supported and funded by FLock.io LTD.
\end{ack}

\bibliographystyle{plain}
\bibliography{references}

\end{document}